\documentclass[12pt]{article}
\input epsf.tex
\def\DESepsf(#1 width #2){\epsfxsize=#2 \epsfbox{#1}}
\usepackage{epsfig}
\usepackage{array}
\makeatletter
\def\alt{\mathrel{\mathpalette\gl@align<}}
\def\agt{\mathrel{\mathpalette\gl@align>}}
\def\gl@align#1#2{\lower.6ex\vbox{\baselineskip\z@skip\lineskip\z@
\ialign{$\m@th#1\hfil##\hfil$\crcr#2\crcr\sim\crcr}}}
\makeatother

\begin{document}
\begin{flushright}
{\tt hep-ph/0503052}\\
March, 2005 
\end{flushright}
\vspace*{2cm}
\begin{center}
{\baselineskip 25pt \Large\bf 
Split Supersymmetry in Unified Models \\
}

\vspace{1cm}

{\large
Bhaskar Dutta and 
Yukihiro Mimura

\vspace{.5cm}

{\it 
Department of Physics, University of Regina, \\
Regina, Saskatchewan S4S 0A2, Canada
}}
\vspace{.5cm}

\vspace{1.5cm}
{\bf Abstract}\end{center}
In the context of split supersymmetry, the gaugino mass spectrum seems to be  very important to satisfy the
dark matter content of the universe and the gauge coupling unification. 
In this paper, we have considered various sources of gaugino masses in the context of unified models. 
We show  that the gaugino mass spectrum varies in different unification pictures. In the context of SU(5), we
have found that the  Bino/Wino mass ratio  can be close to one at the weak scale which is helpful to satisfy the WMAP  data. The gluino/Wino mass ratio is also different from the usual  scenario of unified gaugino masses. 
The gaugino masses can be around one TeV and $m_{\rm SUSY}$ is chosen so that the gluino mass does not create
 any cosmological problem. In the context of the Pati-Salam model, we show that the gluino mass can be made  very heavy even after maintaining the unification of the gauge couplings.

\thispagestyle{empty}

\bigskip
\newpage

\addtocounter{page}{-1}

\section{Introduction}
\baselineskip 18pt

The determination of cosmological parameters \cite{Perlmutter:1998np} 
have  impacts on particle physics.
The revelation of the non-zero cosmological constant (responsible for dark energy) seems to
be  the biggest problem in  particle physics \cite{Weinberg:1988cp}.
It is expected that there should be a mechanism to tune the 
cosmological constant~\cite{Bousso:2000xa}.
Such a mechanism may also be applied to
solve the gauge hierarchy problem~\cite{Silverstein:2004sh}.
In this scenario  the low energy supersymmetry (SUSY) is not a necessity for the
physics beyond the standard model(SM).
In another view point, the 
anthropic landscape of string/M-theory vacua has been discussed recently \cite{Susskind:2003kw},
and the naturalness principle does not have the criterion to select models in such an analysis.
In fact, such a statistical argument of  SUSY breaking vacua
claims that the high energy SUSY breaking is preferred \cite{Susskind:2004uv}
when the cosmological constant is anthropically tuned to be zero and 
there exist multiple sources of SUSY breaking, $m_{3/2}^2 = (\sum_i |F_i|^2 + \sum_i D_i^2)/M_P^2$.
The low energy theories have been discussed 
in the context of such anthropic landscape \cite{Arkani-Hamed:2005yv}.

Under such circumstances, the split SUSY scenario 
\cite{Arkani-Hamed:2004fb,Arvanitaki:2004eu,Giudice:2004tc,Mukhopadhyaya:2004cs,
Zhu:2004ei,Gupta:2004dz,Kokorelis:2004dc}
is suggested as one possibility of a phenomenological model.
In the scenario, the SUSY breaking scale, $m_{\rm SUSY}$,
is large, 
e.g. $10^9$ GeV, 
and the squarks and sleptons are heavy,
while gauginos and Higgsinos remain light (masses $\sim$  weak scale) due to 
$R$-symmetry or other related symmetries. 
Of course, SUSY does not play an essential role to explain the gauge hierarchy, 
but other motivations of supersymmetry are maintained: e.g.
the three gauge couplings of the SM are unified at the GUT scale,
and the lightest neutralino is a candidate to explain the  cold dark matter content of the
Universe.
Since all the sfermions are assumed to be heavy, 
the FCNC related problems are absent and the lightest CP-even neutral Higgs mass
can be large enough compared to the current experimental bound 
\cite{Arkani-Hamed:2004fb,Arvanitaki:2004eu}.
The dangerous proton decay operators are also suppressed.
Therefore, the typical fine-tuning problems of the low energy SUSY models are 
absent,
except for the justification of the fact that one of the Higgs doublets is light \cite{Drees:2005cp}.

In the phenomenological framework of the scenario, 
it is important to choose the SUSY breaking scale.
The scale is written as $m_{\rm SUSY} = F_X/M_P$, where $F_X$ is an $F$-term
of the chiral superfield $X$, 
which is a dominant source of SUSY breaking, and $M_P$ is the Planck scale.
Since the gaugino masses can be assumed to be $m_{\rm SUSY}^2/M_P$ naively (if the
dominant contribution is forbidden by $R$-symmetry),
the typical scale of $m_{\rm SUSY}$ is $10^{10}$ GeV. However, the $m_{\rm SUSY}$ can be $10^6$ GeV (PeV scale) \cite{Wells:2004di} (in this case the 
anomaly mediation seems to  be the dominant source for the gaugino masses),
or can be much larger $\sim 10^{13}$ GeV
(in this case the gauginos are heavy and only Higgsinos masses are light to be at around the TeV scale).
One can also consider the models with  
the Fayet-Iliopoulos $D$-terms \cite{Kors:2004hz}
which generates  another source of SUSY breaking.

\begin{figure}[tbp]
	\includegraphics[width=8cm]{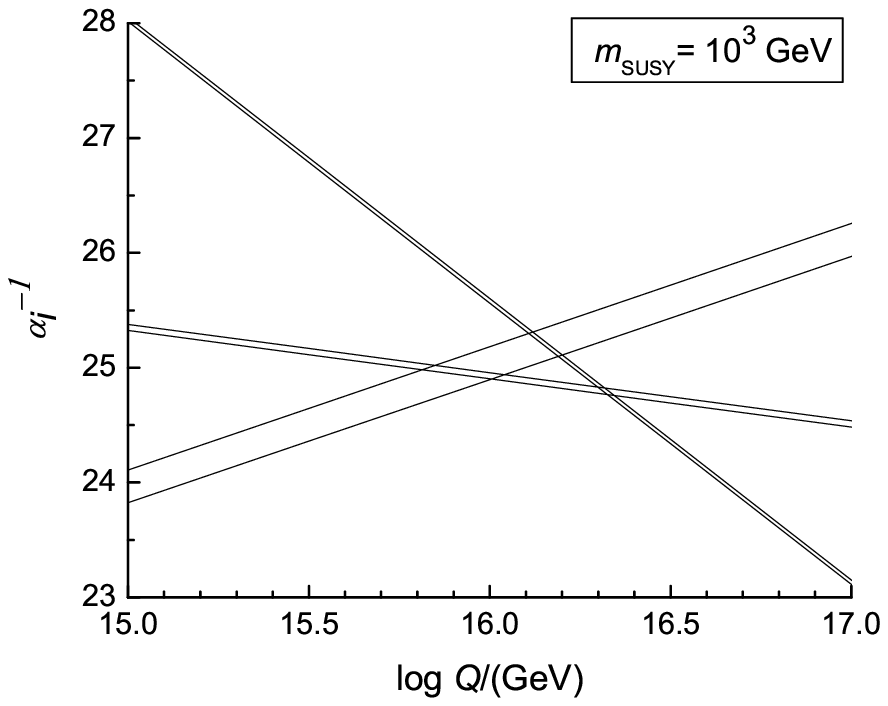}
	\includegraphics[width=8cm]{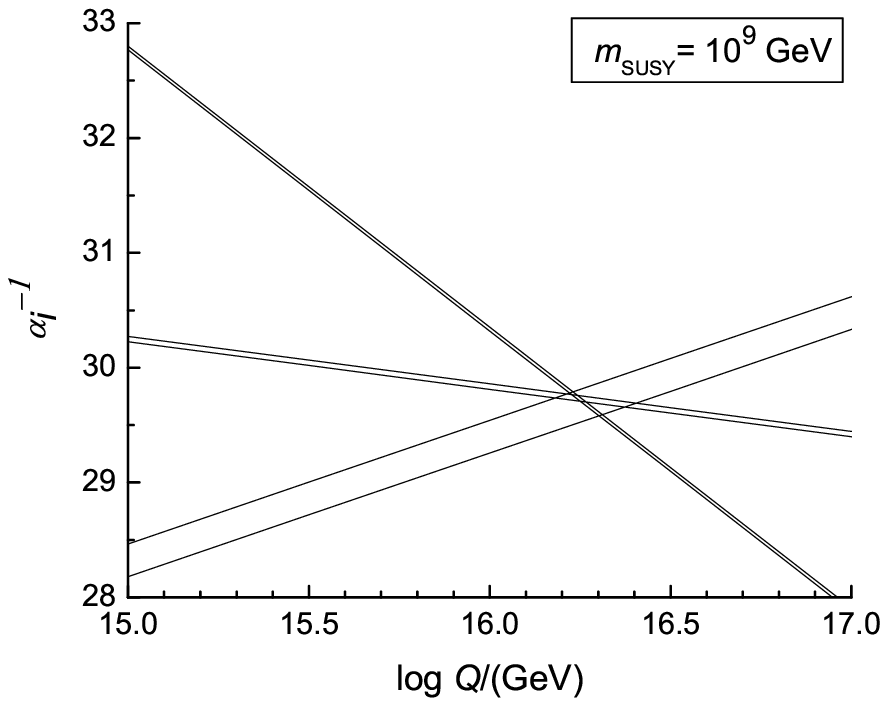}
\caption{Two-loop RGE evolution around the GUT scale in the case $\tan \beta=10$. 
The lines are drawn at 1$\sigma$ range in the case 
$M_3 = \mu = 1$ TeV and $M_2 = 300$ GeV.}
\label{Fig:1}
\end{figure}

Now we will clarify the motivation of our work in the context of split SUSY.
First, let us investigate the gauge coupling unification.
As it is well known, the three gauge couplings are unified at GUT scale
with the following condition at one-loop level,
\begin{equation}
\frac{12}{\alpha_2(M_Z)^{SM}}-\frac{5}{\alpha_1(M_Z)^{SM}}-\frac{7}{\alpha_3(M_Z)^{SM}}
= 
- \frac{19}{4\pi} \ln \frac{M_{SUSY}}{M_Z} \,, \label{RGE-MSSM}
\end{equation}
when all the SUSY particles (and a heavier Higgs doublet) are at a single scale $M_{SUSY}$
and the possible GUT threshold corrections are neglected. 
The $M_{SUSY}$ scale is calculated at around 50 GeV by this one-loop equation, 
but the scale is corrected
to be around one TeV when two-loop contributions are included.
The SUSY threshold correction in the Eq.(\ref{RGE-MSSM}) can be rewritten as a multi-scale threshold,
\begin{equation}
-\frac{19}{4\pi} \ln \frac{M_{SUSY}}{M_Z} \rightarrow
\frac{1}{2\pi} \left(
14 \ln \frac{M_3}{M_2} 
- 2 \ln \frac{M_2}{M_Z} - 6 \ln \frac{\mu}{M_Z} 
+ \frac32 \ln \frac{m_{\tilde q}}{m_{\tilde \ell}} 
- \frac32 \ln \frac{m_H}{M_Z}
\right) \,, \label{RGE-MSSM2}
\end{equation}
where $M_3$, $M_2$ are gluino and Wino masses, 
 $\mu$ is a Higgsino mass,
$m_{\tilde q}$, $m_{\tilde \ell}$ are squark and slepton masses,
and $m_H$ is the  mass of the heavier Higgs doublet.
It is easy to see that the SUSY threshold correction
is insensitive to the scalar masses, $m_{\tilde q}, m_{\tilde \ell}, m_H$,
which we call $m_{\rm SUSY}$ in split SUSY case.
Thus, even if those scalar masses are splitted to be heavy,
the coupling unification can still hold.
It is important to notice that the gluino/Wino mass ratio is a significantly sensitive parameter 
in the Eq.(\ref{RGE-MSSM2}).
In the case of  usual mSUGRA models, the gluino/Wino mass ratio is about 3 at a TeV scale, 
and thus the strong gauge coupling is predicted to be larger at the $M_Z$ scale.
In order  to predict the correct magnitude of the  strong coupling one needs $\mu$ to be $\sim$ 10 TeV, 
but such a direction is disfavored satisfying the WMAP data.
In  split SUSY, $m_H$ can compensate the discrepancy by the choice 
$m_H \sim 10^9$ GeV,
and thus the Higgsino and the Wino can be kept at the TeV scale
and these can be the sources of dark matter.
In Fig.1, we show the 2-loop RGE evolutions of gauge couplings \cite{Jones:1981we} 
in the $\overline{DR}$ scheme
in the case, $M_3 = \mu = 1$ TeV and $M_2 = 300$ GeV.
In the figure, we use $\alpha_3^{\overline{MS}}(M_Z) = 0.1187 \pm 0.002$,
$\sin^2 \theta_W^{\overline{MS}}(M_Z) = 0.23120 \pm 0.00015$, 
and $1/\alpha^{\overline{MS}}(M_Z) = 127.918 \pm 0.018$
\cite{Eidelman:2004wy}.

Next, let us see the neutralino dark matter condition 
to obtain the dark matter density observed as by the WMAP data \cite{Perlmutter:1998np}, 
$0.094 < \Omega_{DM} h^2 < 0.129$.
The Refs.\cite{Giudice:2004tc,Pierce:2004mk} use 
the FORTRAN DarkSUSY package \cite{Gondolo:2002tz}
and find solutions which satisfy WMAP data in the case of split SUSY.
The solutions are on the 2D surface in the $(M_1,M_2,\mu)$ space,
and can be written as the following three typical cases:
\begin{enumerate}
	\item Bino-like dark matter : 
$M_1 \sim M_2 \alt \mu, \ \ {\rm or} \ \ \ M_1 \sim \mu \alt M_2\,$.

	\item Wino dark matter : $M_2 \simeq 2-2.5$ TeV, $ M_2 \sim \mu \alt M_1\,$.

	\item Higgsino dark matter : $\mu \simeq 1$ TeV.
\end{enumerate}
For the Bino dark matter case, the universe is overclosed when $M_1 \ll M_2,\mu$.
So the Bino must be mixed with Wino and/or Higgsino.
When Higgsino is heavier than Bino and Wino, the Bino and Wino masses are
needed to be degenerate, $M_1 \simeq M_2$, to make the mixing large.
In the Bino-like dark matter case above, the LSP neutralino mass should be less than
around 1 TeV.

The gaugino/Higgsino spectrum in split supersymmetry is important to satisfy the WMAP data.
In this paper, we will try to produce the above gaugino spectrum in the context of the unified 
models.
We will see that the cases (1) and (2) can be observed in the context of unified models at the GUT scale
 e.g. $SU(5)$, $SU(4)_c\times SU(2)_L\times SU(2)_R$ etc.

It has been pointed out \cite{Arkani-Hamed:2004fb} that the 
gluino is the most important particle
since its lifetime can be a direct probe of the scale $m_{\rm SUSY}$.
The gluino, in this model, can decay only through the virtual scalar quark and
it is long-lived \cite{Cheung:2004ad} compared to the usual low energy SUSY framework.
The gluino decay rate of $\tilde g \rightarrow q\bar q \chi$ is given as \cite{Djouadi:2000aq}
\begin{equation}
\Gamma \sim \frac{\alpha_s \alpha}{48 \pi} \frac{m_{\tilde g}^5}{m_{\tilde q}^4} (a_{ji}^2 + b_{ji}^2),
\end{equation}
where the masses of quark and neutralino 
for final states are neglected compared
to the gluino mass and $a_{ji},b_{ji}$ are the neutralino-squark-quark couplings which are defined 
in the Ref.\cite{Djouadi:2000aq}.
Since the gluino lifetime is sensitive to the scalar quark mass,
and so the scalar mass scale can be probed from  the lifetime of gluino.
The lifetime of gluino is $1-10^4$ second when the gluino mass is  1 TeV and scalar mass scale is $10^{9-10}$ GeV.
But this will disturb the nucleosynthesis in the early universe.
Since one of the motivation of this framework is also the explanation of the origin of 
cold dark matter, 
it is important to probe the lifetime of gluino.
The lifetime  can determine the allowed ranges of gluino  and scalar quark masses in this framework.
Here, we naively consider two typical cases to make the gluino lifetime
to be around $10^{-7}$ second, which is the freeze-out time of cold dark matter
in the neutralino LSP scenario:
(A) the gluino mass is 1 TeV and the scalar mass scale is $10^7$ GeV,
(B) the gluino mass is heavier to be $10^{4.5}$ GeV and the scalar mass scale is $10^9$ GeV.
The gluino mass scale is the key issue to compare the phenomenological 
models which can be distinguished in the future collider experiments such as LHC and ILC. We will consider
different unified models and show 
that the above conditions for the gluino mass can be  satisfied. We will also show that the gluino mass can be  very heavy in  certain unified models.

In this paper, we discuss several sources of gaugino masses and point out a dominant source in the
case of unified models.
We discuss several unification scenarios and show that the gaugino spectrum is a useful probe of these scenarios.
The Higgsino (for both charged and neutral) mass is characterized 
only by the mass parameter $\mu$,
which is just a model parameter.
It is hard to specify the symmetry that suppresses the $\mu$-term and the reason for one of the Higgs doublets  being light \cite{Drees:2005cp}.
Thus observing a Higgsino is not a good probe to determine the physics at the 
high energy scale.
For gauginos, on the other hand, we have three mass parameters,
and a given gaugino spectrum can be related to a particular unified model. Therefore the observation of the gaugino spectrum can be a good probe
of the theory at a high energy scale.
In section 2, we discuss the origins of gaugino masses,
and point out that the Goldstino contribution may important.  
We discuss the SU(5) unification scenario in section 3, and in
section 4 we discuss SU(4)$_c\,\times$ SU(2)$_L\,\times$ SU(2)$_R$, 
SU(3)$_c\,\times$ SU(2)$_L\,\times$ SU(2)$_R\,\times$ U(1)$_{B-L}$ and SO(10) models   
and we conclude in section 5.

\section{Possible Origins of Gaugino Masses}

As we have seen in the introduction,
the gaugino and the Higgsino mass spectrums are important 
for gauge coupling unification, neutralino dark matter, and also
for the measurements at the colliders.
Since the magnitudes of three gaugino masses are related
to the origin of the gaugino masses, it is worth classifying the origins.
In this section,
we enumerate the possible origins of  gaugino masses.

\subsection{The usual mSUGRA origin of gaugino mass}
The gaugino mass in the superfield formalism is written as
\begin{equation}
\int d^2\theta \left(\frac1{g^2}+ M_{1/2} \theta^2 \right) W^\alpha W_\alpha + H.c. ~,
\end{equation}
where $g$ is the gauge coupling constant. 
In the unified model such as SU(5), the gaugino masses are unified at the GUT scale
just like the gauge coupling constants.
We assume that the contribution from the SUSY breaking, 
$M_{1/2} \sim F_X/M_P \sim m_{\rm SUSY}$,
is forbidden by the $R$-symmetry, and the leading contribution is 
$M_{1/2} \sim F_X F_X^\dagger/M_P^3 \sim m_{\rm SUSY}^2/M_P$.
As usual, the three gaugino masses are unified at the GUT scale also in  the split SUSY scenario.
However, the usual relation in mSUGRA, $M_i = \alpha_i M_{1/2}$, is broken
below $m_{\rm SUSY}$ even at the 1-loop level. 
The RGE evolution below $m_{\rm SUSY}$ is described in Ref.\cite{Giudice:2004tc}.

\subsection{The Anomaly mediation}
Even if the gauge field strength $W_\alpha$ does not couple to the  spurion singlets,
the anomaly mediation \cite{Randall:1998uk} can produce  gaugino masses 
by a chiral compensator field $\phi$ in supergravity,
\begin{equation}
M_i = \frac{\beta(g_i)}{g_i} F_\phi ~. \label{anomaly-mediation}
\end{equation}
If this contribution is dominant as a source for gaugino masses, 
Wino can be the dark matter candidate and $M_2 \simeq 2-2.5$ TeV.
Then
the scalar mass scale is around $10^6$ GeV \cite{Wells:2004di}.
The relation (\ref{anomaly-mediation}) is also modified below $m_{\rm SUSY}$ \cite{Giudice:2004tc}.
Since the gluino/Wino mass ratio is large ($M_3/M_2 \sim 8-9$),
the Higgsino needs to be heavy $\mu \sim 20$ TeV to satisfy the gauge coupling unification.
Thus, every superparticle is heavy in this solution.

\subsection{Gauge mediation}

When the messenger fields $f$,$\bar f$ couple to the  spurion singlet,
\begin{equation}
\int d^2\theta \,(m + \theta^2 F_m) f \bar f + H.c. \,,
\end{equation}
the gauginos masses are produced at the one-loop finite contribution \cite{Dine:1993yw},
\begin{equation}
M_i = \frac{\alpha_i}{4\pi} T_i(f) \,
m \left( \frac{M^2_{f1}}{M_{f1}^2-m^2} \ln \frac{M_{f1}^2}{m^2}
   - \frac{M^2_{f2}}{M_{f2}^2-m^2} \ln \frac{M_{f2}^2}{m^2}   \right) \sin2\beta_m \,,
\end{equation}
where $T_i (f)$ is Dynkin index for $f+\bar f$ and 
$M_{fi}^2$ are the scalar messenger eigenmasses
\begin{equation}
M_{f1,2}^2 = m^2 + \frac{m_f^2 + m_{\bar f}^2}2 \pm \sqrt{|F_m|^2 + 
\left( \frac{m_f^2 - m_{\bar f}^2}2 \right)^2}
\end{equation}
and $\tan 2\beta_m = 2|F_m/(m_f^2-m_{\bar f}^2)|$.
The $m_f^2$, $m_{\bar f}^2$ are the SUSY breaking masses for messenger scalars.

When the messenger scale is much larger than the SUSY breaking scale,
$m^2 \gg F_m, m_{\rm SUSY}^2$,
the usual gauge mediation formula  at the messenger scale $m$ is
\begin{equation}
M_i = \frac{\alpha_i}{4\pi} \, T_i (f) \frac{F_m}{m}\,. \label{gauge-mediation}
\end{equation}
If the messenger scale is much less than $m_{\rm SUSY}$,
the gaugino masses are proportional to $m F_m/m_{\rm SUSY}^2$ instead of $F_m/m \,$.

\subsection{$\mu$-term}

Since the $B$-term is large in the case of split SUSY, the Higgs doublet can work as messenger field.
But, since one of the Higgs scalar is splitted to be light by the fine-tuning, 
$B\mu \sim \mu^2 + m_{\rm SUSY}^2$,
the usual formula of gauge mediation, Eq.(\ref{gauge-mediation}), can not be applied.

%
The wino and the bino masses can be generated from $\mu$ 
by finite Higgsino-Higgs one-loop corrections.
In the case $m_{\rm SUSY} \gg \mu \gg M_W$, we have 
\begin{equation}
M_2 \simeq \frac{\alpha_2}{4\pi}\, \mu\, \sin 2\beta\, \ln \frac{m_{\rm SUSY}^2}{\mu^2} , \quad
M_1 \simeq \frac{\alpha^\prime}{4\pi}\, \mu\, \sin 2\beta\, \ln \frac{m_{\rm SUSY}^2}{\mu^2} \,.
\label{finite-mu}
\end{equation}
We also have one-loop RGE corrections below $m_{\rm SUSY}$ due to the loop diagram 
where the  light Higgs doublet propagates in the loop.
The RGE correction gives rise to same signature in the finite correction in Eq.(\ref{finite-mu}).
Finally, we have a gaugino mass  from $\mu$-term contribution below $\mu$,
\begin{equation}
M_2 \simeq \frac{\alpha_2}{2\pi}\, \mu\, \sin 2\beta\, \ln \frac{m_{\rm SUSY}^2}{\mu^2} , \quad
M_1 \simeq \frac{\alpha^\prime}{2\pi}\, \mu\, \sin 2\beta\, \ln \frac{m_{\rm SUSY}^2}{\mu^2} \,.
\label{mu-contribution}
\end{equation}

However, this contribution (just by itself) is not favored if neutralino is the dark matter
candidate,
since the bino nature of the LSP may overclose the universe.
We need another source of gaugino masses to make $M_2 \sim M_1$ or $M_2 \ll M_1$
in the case $M_{1,2} \ll \mu$.
The possibility of heavy Higgsino is studied in Ref.\cite{Cheung:2005ba}.

\subsection{The $A$-term}
\subsubsection{The scalar trilinear couplings for Sfermions}
The Sfermion trilinear couplings ($A_f \tilde q \tilde u^c h_u$) generates gaugino masses 
              through two-loop RGE \cite{Yamada:1993ga}
above $m_{\rm SUSY}$.

Since the $A_f$ term breaks $R$-symmetry, one can forbid this contribution
in the split SUSY scenario
in the same way as original gaugino masses are forbidden.
However, if one protects  the gaugino (and/or Higgsino) masses 
by an anomalous $U(1)$ symmetry, but breaks $R$-symmetry,
this contribution can be dominant \cite{Babu:2005ui}.

When only the top Yukawa contribution is dominant, the gaugino masses at the SUSY breaking scale 
are  approximately:
\begin{equation}
M_i \sim 2 c_i \frac{\alpha_i \alpha_t}{(4\pi)^2} A_t \ln \frac{M_G}{m_{\rm SUSY}},
\end{equation}
where $\alpha_t=y_t^2/(4\pi)$, $c_i = (26/5,6,4)$ and $A_t$ is the $A$-parameter for 
top-stop-Higgs coupling.
In this case, the Bino is the lightest  among the gauginos at the weak scale.
So  when this contribution is
dominant,  the Higgsino mass needs to be chosen appropriately to satisfy the WMAP data.
\subsubsection{The Goldstino contribution}

\begin{figure}[tbp]
\centering
	\includegraphics[width=6cm]{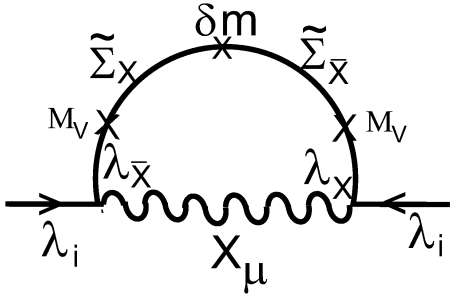}
\caption{One-loop diagram which gives the finite correction to the gaugino masses.}
\label{Goldstino-diagram}
\end{figure}

In the super Higgs mechanism, the Goldstino $\widetilde\Sigma$ 
is the Higgsino eaten by the heavy gaugino.
The Goldstinos acquire masses at the symmetry breaking scale by 
folding the Dirac mass with heavy gauginos, $\lambda_X$,
corresponding to the broken generators of gauge symmetry.
In the SUSY limit, the bilinear terms of Goldstinos are absent
since their superpartners (the would-be Goldstone bosons) are massless.
If the SUSY is broken in the GUT Higgs potential, the
Goldstino bilinear term is non-zero, 
\begin{equation}
\delta m \sim A_{\rm GUT} + \frac{m_{\rm SUSY}^2}{M_{\rm GUT}},
\label{Goldstino-mass}
\end{equation}
where $A_{\rm GUT}$ is a dimension one SUSY breaking parameter 
in the GUT Higgs potential.
Then the heavy gauginos become pseudo-Dirac fields with following the mass terms
\begin{equation}
(\lambda_X, \widetilde\Sigma_X)  
\left( \begin{array}{cc} M_{1/2} & M_V  \\ -M_V & \delta m    \end{array} \right)  
\left(\begin{array}{c} \lambda_{\bar X} \\ \widetilde\Sigma_{\bar X}  \end{array}  \right) + H.c. ,
\end{equation}
and the heavy gaugino 
loop diagram, as shown in Fig.\ref{Goldstino-diagram}, generates the finite corrections to the masses 
for the SM gauginos \cite{Hisano:1993zu},
\begin{equation}
M_i = \frac{\alpha_G}{2\pi} \, T_i(\Sigma_X) \, \delta m \,.
\label{Goldstino-mass1}
\end{equation}
We also have the contribution which is proportional to $M_{1/2}$
as a one-loop correction to the  original gaugino mass $M_{1/2}$.

The MSSM superpotential can be invariant under the  $R$-symmetry and thus the $A_f$ 
contribution can be always controlled by $R$-symmetry.
The other contributions listed above also can be forbidden by $R$-symmetry.
On the other hand,
the realistic GUT Higgs superpotential itself breaks $R$-symmetry,
and thus the GUT symmetry breaking always causes the $R$-symmetry breaking
and contributes to the gaugino masses through the Goldstino loop.
%
Consequently, this may introduce a dominant contribution to the gaugino mass
in the split SUSY scenario
as long as the unified symmetry is broken by Higgs mechanism,  
unless we introduce  some other symmetries to control it. 

Note that the $m_{\rm SUSY}$ in Eq.(\ref{Goldstino-mass}) 
is the scalar mass for GUT Higgs fields.
 In order to keep  the gaugino masses less than 1 TeV, we need $m_{\rm SUSY} \alt 10^{11}$ GeV
when $M_{\rm GUT} \sim 10^{16}$ GeV.


\subsection{Selection of the origins}

All these above contributions can be present in the gaugino masses
and any one of the above sources can generate 
 dominant contribution.
The parameter space of $(M_1,M_2,\mu)$ is restricted by phenomenologies. 
For example, if the Higgsinos are much heavier than the gauginos, e.g. $\mu \simeq 10$ TeV,
we need $M_2 \alt M_1$ and the wino-like neutralino is the LSP so that we do not exceed 
the amount of relic density \cite{Pierce:2004mk}.
Therefore, in the heavy Higgsino case, the usual mSUGRA source is not favored very much and
the anomaly mediation is a good candidate.
In the case of anomaly mediation, $M_2 \sim 2-2.5$ TeV
is required for the  Wino dark matter. The gravitino and the scalar masses are 
about $10^6$ TeV \cite{Wells:2004di}. 
If the  Higgsinos are light, 
the Higgsino-like neutralino can be the LSP.
However, in this case, the masses of lightest neutralino and chargino are
degenerate if both $M_1$ and $M_2$ are large and the relic density can be very small
due to the coannihilation processes.
In order to satisfy the WMAP data, the Higgsino should be around 1 TeV \cite{Pierce:2004mk}.


If the origin of gaugino masses produces the three gaugino mass parameters 
of the same order, 
the gluino mass is less than several TeV (since the dark matter constraint require the other gaugino
masses to be $\sim$ 1 TeV).
This means that the lower scale of SUSY breaking is favored
due to the lifetime of the gluino as we have discussed in the introduction.
So, the anomaly mediation with the PeV scale $m_{\rm SUSY}$ may be a good scenario.
However, all the superparticles need to be heavier than 2 TeV
and such a situation is not very attractive for the precision electroweak data.
On the other hand, 
if the origin of the gauginos masses is due to the Goldstino 
contribution, which happens in the case of the unifying models, 
the gaugino spectrum  depends on the unification group. 
The precise dependence of the spectrum  on  different  unifying gauge symmetries 
are  discussed in  the next two sections.

\section{SU(5)}


The $A$-term in the GUT Higgs potential as shown in the Eq.(\ref{Goldstino-mass}) can be 
a dominant contribution of gaugino masses in split SUSY scenario.
In this section, we study the contribution in the SU(5) GUT model.
The SU(5) symmetry is broken down to SM gauge group by the
VEV of adjoint Higgs field $\Sigma$.
The adjoint Higgs field is decomposed as
\begin{equation}
\Sigma: {\bf 24} = \Sigma_8 : ({\bf 8}, {\bf1})_0 + \Sigma_3 : ({\bf 1}, {\bf3})_0
+ \Sigma_1 : ({\bf 1}, {\bf1})_0 + \Sigma_X : ({\bf 3}, {\bf2})_{-5/6}
+ \Sigma_{\bar X} : (\overline{\bf 3}, {\bf2})_{5/6} \,.
\end{equation}
%
When the singlet $\Sigma_1$ get VEV ($\langle \Sigma_1 \rangle = \sigma$),
the would-be Goldstone bosons, $\Sigma_{X,\bar X}$,
are eaten by heavy gauge bosons,
and their superpartners, Goldstinos $\tilde\Sigma_{X,\bar X}$, combine Dirac masses with 
heavy gauginos, $\lambda_{X,\bar X}$.

The Goldstino bilinear mass term is 
$(M-\lambda \sigma/\sqrt{30})\, \tilde\Sigma_X \tilde\Sigma_{\bar X}\,$ 
for the Higgs superpotential which breaks SU(5) symmetry 
\begin{equation}
W = \frac{M}2 \,{\rm Tr}\, \Sigma^2 + \frac{\lambda}3 \,{\rm Tr}\, \Sigma^3 \,. \label{GUT-W}
\end{equation}

%
In the SUSY limit, the VEV is $\sigma = \sigma_0= \sqrt{30} M/\lambda$,
and
the bilinear term of the Goldstino is of course absent.
However, if we include the soft SUSY breaking terms
\begin{equation}
V_{\rm soft} = m_\Sigma^2 \,{\rm Tr}\, \Sigma \Sigma^\dag
+ \left( \frac12 B_G M \,{\rm Tr}\, \Sigma^2 + \frac13 \lambda A_G \,{\rm Tr}\, \Sigma^3 
+ H.c. \right)\,,
\end{equation}
the VEV shifts to $\sigma_0 +\Delta \sigma$.
Then the Goldstino bilinear mass $\delta m$ arises
\begin{equation}
\delta m = -\frac{\lambda}{\sqrt{30}} \Delta \sigma \simeq B_G - A_G + \frac{m_\Sigma^2}{M} \,,
\end{equation}
%
%
and the one-loop diagram in which heavy gauge bosons and gauginos propagate in Fig.2
induces the gaugino masses $M_i$ at GUT scale,
\begin{equation}
M_i = \frac{\alpha_G}{2\pi} \, T_i(\Sigma_X) \, \delta m \,.
\end{equation}
Since the heavy multiplet, $({\bf 3}, {\bf2})_{-5/6}+(\overline{\bf 3}, {\bf2})_{5/6}\,$,
 are non-singlets under each gauge symmetry,
it contributes to all gaugino masses $M_i$,
and the ratios at the GUT scale are
\begin{equation}
M_1 : M_2 : M_3 = 5 : 3 : 2 \,. \label{gaugino-mass-ratio}
\end{equation}
Since this ratio is different from the original unified gaugino source or anomaly
mediation,
one can distinguish this scenario from the measurement of the gaugino mass spectrum.

We note that the Goldstino contribution is not related to the details of the GUT superpotential.
The gaugino mass ratios depends only on the quantum numbers of the broken generators.
So even if we employ $\bf 75$ Higgs to break SU(5) down to the SM,
the mass ratio (\ref{gaugino-mass-ratio}) does not change
as long as the unified group is SU(5).

Let us see how the Goldstino contribution can dominate over the others.
We assume that the spurion fields $X$ is the dominant source of the SUSY breaking.
We assign a $R$-charge $-1$ to other spurion $S$ field and 1 to $\Sigma$,
so that the superpotential 
$W= M \Sigma^2/2 + S \Sigma^3/M_P$ can be $R$-symmetric.
We assign a positive and fractional $R$-charge to the spurion $X$ in order 
not to disturb the superpotential.
Then the coupling $\lambda$ in Eq.(\ref{GUT-W}) is
$\lambda = 3S/M_P$ 
and the $A$ parameter is 
$A_G = F_S/S$, and $A_G$ is a free parameter ($A_G < m_{\rm SUSY}$ by assumption).
We can suppress all other sources of gaugino masses
as is done in the case of split SUSY. 
For example, if the gluino mass is around a TeV from the Goldstino contribution, 
$m_{\rm SUSY}$ is around $10^7$ GeV  
for the  cosmological reason. 
The usual mSUGRA leading contribution $M_i \sim m_{\rm SUSY}$ is forbidden by the $R$-symmetry,
and thus the direct SUGRA contribution is $m_{\rm SUSY}^2/M_P\sim 10^{-4}$ GeV.
Since the $R$-breaking is incorporated in the GUT symmetry breaking,
the dimension one contribution from the SUSY breaking spurion $S$ with the one-loop factor
can dominate the gaugino masses.

In this example, we have used $R$-symmetry to control the superpotential. 
Note that if the $R$-symmetry is the exact symmetry of the Lagrangian, $the R$-axion 
may cause the cosmological problem.
Since we are not specifying the hidden sector potential, this concern is beyond the scope of the present paper.
However, if the axion-like problem is considered seriously, 
one can construct models by using anomalous U(1) gauge symmetry in a similar way.

Note that if $B_G$ is large, the
heavy chiral multiplets such as $\Sigma_8$ and $\Sigma_3$
work as messengers and they contribute to gaugino masses.
In the above example of $R$-charge assignment,
the magnitude of the $B$-term is
$B_G M \sim$ $m_{\rm SUSY}^2 (S/M_P)^2$,
and thus $B_G$ itself is at most around weak scale in the split SUSY context
and the messenger contribution is not very large.
If we also make the adjoint Higgs mass $M$ to be the spurion mass and reassign $R$-charges appropriately,
$B_G$ becomes free parameter $\sim F_M/M$ and 
such messenger effects can contribute to the gaugino masses. 
This $B$-term contribution depends on the details of the GUT particle spectrum and if this contribution dominates the
gaugino mass spectrum will not exhibit the pattern of Eq.(\ref{gaugino-mass-ratio}).

\begin{figure}[tbp]
\centering
	\includegraphics[width=8cm]{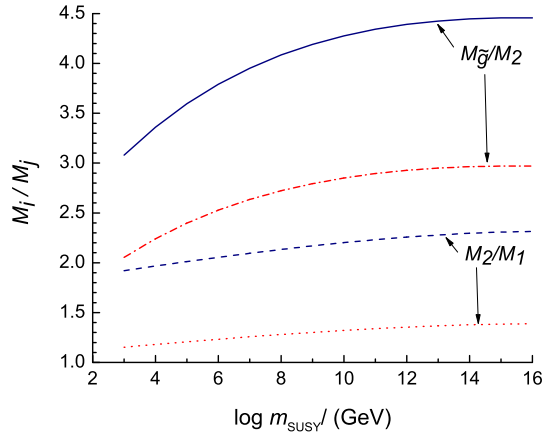}
\caption{Gaugino mass ratios for the unified gaugino case (dark blue lines)
and for the case that Goldstino contribution dominates (red lines).
The values are given in the case where gluino mass is 1 TeV.}
\label{gutsu5}
\end{figure}

We now show the gaugino mass ratios when the gluino mass is 1 TeV
in Fig.\ref{gutsu5}.
The ratio depends on the SUSY breaking scale since the relation ($M_i/\alpha_i=$ const.)
is broken even at a one-loop RGE below $m_{\rm SUSY}$.
The dark blue lines are for the usual ratio $M_1:M_2:M_3=1:1:1$ at GUT scale
and red lines are for the Goldstino contribution $M_1:M_2:M_3 = 5:3:2$.
The gluino mass is shown as a pole mass in $\overline{DR}$ scheme.
%
It is interesting that the Wino/Bino mass ratio $M_2/M_1$ is close to one in the case
of dominating Goldstino contribution.
This is favored as a solution of Bino-like dark matter to satisfy the WMAP data.
In this Bino-like dark matter with $M_1 \sim M_2$, 
the condition is $100 \ {\rm GeV} \alt M_1 \alt 1$ TeV and 
$\mu \agt M_1$.
In the case of the  Wino and the Higgsino type dark matter, 
the Wino and the Higgsino masses  are larger than 1 TeV.
Thus the  Bino-like dark matter prediction due to  the SU(5) Goldstino contributions is attractive,
since  the dark matter candidate and the NLSP can be light and therefore, can be measured at the LHC.


If the Bino/Wino mass ratio is around one, the mass difference ($\Delta M$) between the 
$\tilde\chi^{\pm}_1$  and the $\tilde\chi^0_1$  is very small (the precise value depends on $\mu$). 
Depending on the mass
difference, the final states of a $\tilde\chi^{\pm}_1$ decay will contain lepton plus missing energy 
(arising from $\nu$ and
$\tilde\chi^0_1$) and
$q q'$ plus missing energy (for very small $\Delta M$, there will be pions)  or stable tracks \cite{guchen}. 
The final state particles are soft. An accurate mass measurement  becomes 
difficult. 
 In the case of $\tilde\chi^0_2$,  the dependence of $\mu$ is also quite significant. 
The mass difference between the 
$\tilde\chi^{0}_2$  and the $\tilde\chi^0_1$ depends on $\mu$ crucially.
 We will have
$l\bar l$ plus missing energy and
$q \bar q$ plus missing energy in the final states and depending on the mass difference between 
 $\tilde\chi^0_1$ and
$\tilde\chi^0_2$, the final state particles could be very low in energy.
As before, an accurate mass measurement  becomes 
difficult. It is  also possible that the $\tilde\chi^0_2$ may not decay inside the detectors (if $\mu$ is quite
large). It is interesting to investigate all these final states at
LHC \cite{later}.

\section{Other Unification Groups}

The gaugino mass spectrum from the Goldstino contributions 
does not depend on the details of the Higgs potential,
but it rather depends on the 
quantum number of the broken generators for the Goldstino contribution.
Therefore, if this Goldstino contribution dominates,
the gaugino mass spectrum can be a good probe of how the SM gauge group is unified.

In the SU(5) Goldstino contribution, the 
broken generators corresponds to $({\bf 3},{\bf2})_{-5/6}+(\overline{\bf 3},{\bf 2})_{5/6}$
and thus all three gauginos acquire  masses at the same order.
So, as long as the neutralino relic density dominates
dark matter,
the gluino mass should be around one TeV,
and the SUSY breaking scale will be at the PeV scale
to make the gluino lifetime shorter.

In this section,
we will study the partial unification group of the SM and other possible unification scenarios.
The partial unification groups are
$G_{3221} = {\rm SU(3)}_c \times {\rm SU(2)}_L \times {\rm SU(2)}_R \times {\rm U(1)}_{B-L}\, $,
and $G_{422} = {\rm SU(4)}_c \times {\rm SU(2)}_L \times {\rm SU(2)}_R\, $.

In the case of $G_{3221}$, 
the Goldstino which is non-singlet under the SM is $({\bf 1},{\bf 1})_{\pm 1}$.
Thus only the  Bino mass $M_1$ is influenced by the Goldstino contribution and hence it is heavy.
In this case, the neutralino dark matter can be either the Wino or the Higgsino type.
Since the $G_{3221}$ breaking scale $M_{3221}$ can be much smaller than GUT scale,
the Goldstino bilinear mass can be large even if $A,B$-terms are small,
$\delta m \sim m_{\rm SUSY}^2/M_{2231}$.

In the case of $G_{422}$,
the non-singlet Goldstinos are $({\bf 3},{\bf 1})_{2/3}+(\overline{\bf 3},{\bf 1})_{-2/3}$
in addition to $({\bf 1},{\bf 1})_{\pm 1}$.
Therefore, the 
$G_{422}$ symmetry breaking contributes to both gluino and Bino masses and hence they are heavy.
But the Wino mass does not have this contribution and hence can be light. 
This situation is suitable for the Wino LSP.
Since the gluino can be heavy, the $m_{\rm SUSY}$ can be large $\sim$ $10^9$ GeV
even if  the lifetime of the gluino is constrained.
For example, we can consider a hierarchical spectrum,
$M_3 \sim 10^8$ GeV, $M_1 \sim 10^7$ GeV, $\mu \sim 10^4$ GeV, and $M_2$ can be $\sim$ TeV scale.
We note that $\mu$ is generated from $M_1$ through the one-loop RGE below $m_{\rm SUSY}$,
and $M_2$ can be generated from $M_3$ at the 3-loop order by the RGE effect above $m_{\rm SUSY}$.

\begin{figure}[tbp]
\centering
	\includegraphics[width=8cm]{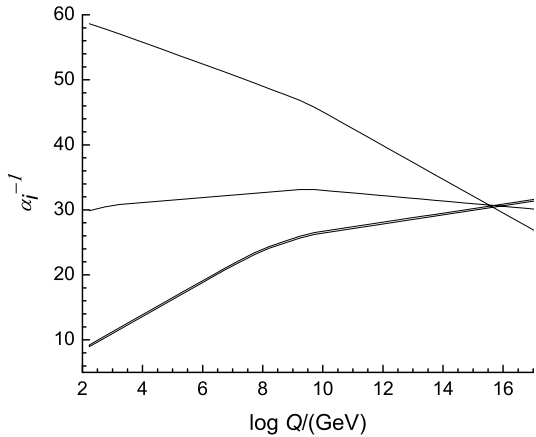}
\caption{Gauge coupling running (two-loop) in the case where the $G_{422}$ Goldstino contributions are large.
The gluino mass is assumed to be $10^8$ GeV. The colored Higgs fields are  added to generate 
the unification.}
\label{PS}
\end{figure}

The gauge coupling unification is not good when $M_3 \gg M_2$.
However we can use the  light colored Higgs fields 
($({\bf 3},{\bf 1})_{1/3}+(\overline{\bf 3},{\bf 1})_{-1/3}$) to help unify the gauge couplings.
Since $m_{\rm SUSY}$ can be $\sim 10^9$ GeV, the proton decay operator
mediated by the colored Higgsino exchange is not dangerous.
The colored Higgsino masses can be generated by $M_3$ at 1-loop  $\sim 10 ^7$ GeV and
the three gauge couplings are unified as shown in Fig.\ref{PS}. 
The gauge couplings unify at $(3-6) \times 10^{15}$ GeV. 

In the case of flipped SU(5), the gaugino mass ratios at the GUT scale are
\begin{equation}
M_1 : M_2 : M_3 = 1/5 : 3 : 2 \,. \label{gaugino-mass-ratio1}
\end{equation} 
So the ratio of the Bino mass to the Wino mass at the weak scale is 1:30. The Bino mass 
is the lightest and this
situation is not good for dark matter unless the value of $\mu$ is around the bino mass. 
The gluino/Wino mass ratio however is same as in the SU(5) case.

The partial unification groups can be unified to SO(10).
The broken generators in the SO(10) are 
$({\bf 3},{\bf2})_{-5/6}+({\bf 3},{\bf2})_{1/6}+(\overline{\bf 3},{\bf 1})_{-2/3}+({\bf 1},{\bf 1})_{1} + c.c.$
using the SM quantum numbers. The first one corresponds to the SU(5) broken generator and
the second one corresponds to the flipped-SU(5) broken generator.
The other terms correspond to the $G_{422}$. 
Each broken generator can contribute to the gaugino mass spectrum.
Depending on the Goldstino spectrum, each contribution can contribute coherently  or some of the 
contributions are dominant. For example, 
in the limit that each multiplet contributes equally,
we find the ratio at the GUT scale:
$M_1 : M_2 : M_3 = 8 : 6 : 5$, and again it is different from the unified scenario.
It is interesting that the measurement of the  gaugino mass spectrum  at the colliders will tell us about
 the  broken generators.

\section{Conclusion}

The gaugino and the Higgsino masses are important for the motivations of SUSY which accommodates
the gauge coupling unification and the neutralino dark matter.
In the split SUSY scenario,
it is assumed that
only the gauginos and the Higgsinos are splitted to be light
and their masses at the low energy scale are important  to realize the motivation of SUSY.

In this paper, we have considered various sources of gaugino masses.
Among them, we focus on the Goldstino contributions  which arise in the case of unified
gauge symmetry models.
In the usual supergravity models, the  unified gaugino mass at GUT scale
is the  leading contribution,
and the Goldstino contribution is a higher order correction.
However, if the original unified gaugino masses  suppressed to be 
$F_X F_X^\dagger/M_P^3$
by $R$-symmetry (or other related symmetries),
the Goldstino contribution may dominate.
This is because the GUT superpotential usually breaks the $R$-symmetry.
The GUT symmetry breaking requires the $R$-symmetry breaking,
and generates a gaugino mass $\sim F_X/X$ with one-loop factor.
The Goldstino contribution can dominate, as long as we consider the unified models  
in the split SUSY scenario with gauge symmetry being broken by the Higgs mechanism.

Interestingly,
the Goldstino contribution depends only on the quantum numbers
of the broken generators.
Thus the gaugino mass spectrum is predictive especially in the SU(5) case.
As we have shown in Fig.\ref{gutsu5},
the Wino and Bino masses are almost same $M_1 \simeq M_2$
contrary to the mSUGRA case, $M_2/M_1 \simeq 2$ at the weak scale.
This is favored for the Bino-like dark matter in the split SUSY model to satisfy the recent WMAP data.
The other dark matter solutions, the Higgsino or the Wino dark matter scenarios
require the LSP mass $\agt$ 1 TeV, whereas the Bino-like LSP mass can be less than a 1 TeV.
Thus this prediction is attractive for the precision electroweak data
and for the future measurements at LHC. Further,
in this scenario, the gluino does not create any problem in the cosmological context since 
the SUSY breaking scale can be $\alt 10^7$ GeV. In the scenarios where the Bino and the Wino mass are very close
or Wino is the LSP, we have soft jets, 
leptons in the final states of $\tilde\chi^{\pm}$, $\tilde\chi^0_2$ decays.

We also considered other partial unification groups.
In the case of $G_{422}$, 
only the Wino mass can be light among the three gauginos.
In this case, the gluino can be heavy and harmless even 
if the SUSY breaking scale is large.
In this scenario, the WMAP data is satisfied by the Wino dark matter.

\section*{Acknowledgments}

This work is supported by 
the Natural Sciences and Engineering Research Council of Canada.

\end{document}